\documentstyle[twocolumn,aps]{revtex}
\begin{document}
\input psfig
\draft

%
\twocolumn[\hsize\textwidth\columnwidth\hsize\csname@twocolumnfalse\endcsname
%
%

\title{Fermi Surface and Spectral Functions 
\\ of a Hole Doped Spin-Fermion Model for Cuprates}

\author{Mohammad Moraghebi$^1$, Charles Buhler$^1$, Seiji Yunoki$^2$ 
and Adriana Moreo$^1$}

\address{$^1$Department of Physics, National High Magnetic Field Lab and
MARTECH,\\ Florida State University, Tallahassee, FL 32306, USA}

\address{$^2$Materials Science Center, \\ University of Groningen,
Nijenborgh 4, 9747 AG Groningen, The Netherlands}

\date{\today}
\maketitle

\begin{abstract}

Using numerical techniques we study the spectral function $A(k,\omega)$
of a spin-fermion model for cuprates in the regime where magnetic and charge
domains (stripes) are developed upon hole-doping. 
>From $A(k,\omega)$ we study the
electronic dynamics and 
determine the Fermi Surface (FS), which is compared with angular resolved
photoemission results for $\rm La_{2-x}Sr_xCuO_2$. A pseudogap is
observed in the density of states at the chemical potential for all
finite dopings. The striped ground state
appears to be metallic in this model 
since there is finite spectral weight at the chemical potential, 
but the electronic hopping seems to be stronger
perpendicular to the stripes rather than along them. 
The band structure is not rigid, contrary to the behavior found 
in mean-field studies, 
and changes with doping. Both mid-gap (stripe induced) 
and valence band states determine the FS.
For vertical (horizontal) stripes, a clear FS appears close to
$(\pi,0)$ $((0,\pi))$, while no FS is observed close to $(0,\pi)$
($(\pi,0)$). Along the diagonal direction the spectral function shows a
clear quasi-particle peak close to (0,0), but its weight is reduced
as the chemical potential is approached. A weak FS
develops along this direction as the system is doped.

\end{abstract}

\vskip2pc]
\narrowtext

\section{Introduction}

Neutron scattering studies have shown that many high temperature
superconducting  materials 
exhibit magnetic
incommensurability (IC) upon hole doping.\cite {Cheong}
In some compounds it is believed that this spin IC
is due to the formation of charge stripes.\cite{Tran} Indirect evidence
of static stripes formation has been observed in Nd doped 
$\rm La_{2-x}Sr_xCuO_2$ (LSCO).\cite{Tran} On the other hand, 
in LSCO, angular resolved photoemission experiments (ARPES)\cite{Ino} 
as well as
neutron scattering\cite{Billinge}, show features consistent with the 
existence of dynamical stripes. In neutron scattering the evidence is in the
broadening of the peak in the supposedly inhomogeneous stripe 
regime, while in ARPES a 
one-dimensional-like band dispersion and a straight Fermi surface (FS) 
around $(\pi,0)$ and $(0,\pi)$ are observed.

The goal of this paper is to gain theoretical understanding on the effect of
stripes in the spectral function and FS of a system of electrons and
spins, since these are
properties which can be measured in ARPES
experiments. In previous attempts to study these properties, 
mean-field and exact diagonalization (ED) techniques have been 
applied to the Hubbard and t-J 
models.\cite{Ichioka,Sadamichi} The problem with these models is that
it has not been shown that their actual ground-state has striped
characteristics.
In mean field approaches
\cite{Ichioka} not all possible states have been considered, and 
with ED, additional attractive terms have to be added to the Hamiltonian 
to stabilize the striped phase.\cite{Sadamichi} 
Thus, here we consider a
spin-fermion model which can be unbiasedly 
studied with accurate Monte Carlo (MC) techniques
without "sign problems" and which, 
upon hole-doping, has a striped ground-state, as shown in previous
investigations.\cite{Charlie,Charlie1}
This model has been used successfully to qualitatively describe magnetic 
and charge properties of the cuprates, showing that spin IC and charge
stripe formation are related.\cite{Charlie,Charlie1}

The main result of the present effort 
is that mid-gap states, associated with the stripes, and 
valence band states related to the background, both contribute to 
determine the FS. 
The band structure is not rigid, as found in mean-field calculations.
\cite{Ichioka} A pseudogap at the chemical potential clearly develops 
as the system is doped. The doped
holes are not only introduced into the stripes but also into 
the background, and the striped state appears to be metallic, although
the electronic hopping is stronger along the direction
perpendicular to the stripes.

The paper is organized as follows: the spin-fermion model is presented
in section II, an schematic picture is shown in section III, and section
IV is devoted to the results. The pseudogap is discussed in section V,
and the conclusions are presented in section VI.

\section{The Model}

The spin-fermion model is constructed as an interacting system of
electrons and spins, crudely mimicking phenomenologically the
coexistence of charge and spin degrees of freedom in 
the cuprates.\cite{Pines,Schrieffer,Fedro}. Its Hamiltonian is given by
$$
{\rm H=
-t{ \sum_{\langle {\bf ij} \rangle\alpha}(c^{\dagger}_{{\bf i}\alpha}
c_{{\bf j}\alpha}+h.c.)}}
+{\rm J
\sum_{{\bf i}}
{\bf{s}}_{\bf i}\cdot{\bf{S}}_{\bf i}
+J'\sum_{\langle {\bf ij} \rangle}{\bf{S}}_{\bf i} \cdot{\bf{S}}_{\bf j}},
\eqno(1)
$$
\noindent where ${\rm c^{\dagger}_{{\bf i}\alpha} }$ creates an electron
at site ${\bf i}=({\rm i_x,i_y})$ with spin projection $\alpha$,  
${\bf s_i}$=$\rm \sum_{\alpha\beta} 
c^{\dagger}_{{\bf i}\alpha}{\bf{\sigma}}_{\alpha\beta}c_{{\bf
i}\beta}$ is the spin of the mobile electron, the  Pauli
matrices are denoted by ${\bf{\sigma}}$,
${\bf{S}_i}$ is the localized
spin at site ${\bf i}$,
${ \langle {\bf ij} \rangle }$ denotes nearest-neighbor (NN)
lattice sites,
${\rm t}$ is the NN-hopping amplitude for the electrons,
${\rm J>0}$ is an antiferromagnetic (AF) coupling between the spins of
the mobile and localized degrees of freedom at the same site,
and ${\rm J'>0}$ is a direct AF coupling
between the localized spins in nearest neighbor sites.
The density $\langle n \rangle$=$\rm 1-x$ of 
itinerant electrons is controlled by a chemical potential $\mu$. 
Hereafter ${\rm t=1}$ will be used as the unit of energy. 
>From 
previous phenomenological analysis the coupling ${\rm J}$ 
is expected to be larger than t, while the Heisenberg coupling
${\rm J'}$ is expected to be smaller.\cite{Schrieffer,Fedro,Charlie} 
The value of ${\rm J}$ will be here fixed to 2 and
the coupling ${\rm J'=0.05}$, as in Ref.\cite{Charlie}.
To simplify the numerical calculations, avoiding the sign problem, the
localized spins are assumed to be classical (with $\rm |S_{\bf i}|$=1).
This approximation was discussed in 
detail in Ref. \cite{Charlie} and it is not expected to alter
qualitatively the behavior of the striped ground-state. The model will
be studied using a standard
MC method, details of which can be found in Ref.~\cite{yuno}.
Periodic boundary conditions (PBC) are  
used.

The spectral function $A(k,\omega)$ is defined as
$$
A(k,\omega)=-{1\over{\pi}}Im G(k,\omega),
\eqno(2)
$$
\noindent where $G(k,\omega)$ is the one-particle retarded
Green's function for the
electrons, $k$ is the momentum and $\omega$ denotes the energy. 
Since we are performing a MC calculation
on the classical spins only, because the fermions do not 
interact among themselves directly but only through the classical
spins, the time-dependent Green's function can be
straightforwardly calculated in real-time.
The spectral functions 
obtained with this procedure
only contain small controlled statistical errors. This should be
contrasted against other calculations that use the 
Maximum Entropy technique which is typically an uncontrolled procedure. 
The density of states (DOS) $N(\omega)$ is obtained
by adding the spectral functions for all different momenta in the first
Brillouin zone (FBZ).

\section{Schematic Picture}

ARPES results 
in ${\rm La_{1.28} Nd_{0.6} Sr_{0.12} CuO_4}$ show
a FS which appears to be formed by the superposition of two
one-dimensional (1D) FS believed to be caused by metallic stripes in an 
insulating
background.\cite{Zhou} Similar features are observed in underdoped LSCO
\cite{Ino} and in ${\rm Bi_2Sr_2CaCu_2O_{8+y}}$ (Bi2212). \cite{Feng}
The naive band structure which would provide such a FS is
shown in Fig.1-a, along the main directions in the FBZ, 
for the case of quarter-filled (metallic) equally-spaced vertical 
stripes, with a period a=4, 
on an AF (insulating) background. The overall density is
$\langle n \rangle=0.875$ which corresponds to a hole density $x=0.125$. 
A valence and a conduction
band, separated by the AF gap, are associated with the insulating
background. The dashed line indicates "shadow bands".\cite{schrief} In
the middle of the gap there is a band that corresponds to the 1D
stripes. Notice that the mid-gap band is not symmetric under $\pi/2$
rotations. At $(\pi,0)$ [$(0,\pi)$] the mid-gap band is closer to the
valence [conduction] band.
The mid-gap band energy is given by
$E_m(k_x,k_y)=t'cos k_y$, while the energy of the valence and conduction
bands are given by $E_k^{\pm}=\pm\sqrt{\epsilon_k^2+\Delta^2}$ where
$\epsilon_k=-2t(cos k_x+cos k_y)$, and $2\Delta$ is the AF gap. In
Fig.1-a we have used $t'=0.5$, $t=1$ and $\Delta=1$ for simplicity.
The metallic band is crossed by the Fermi energy ($E_F$)
defining the FS shown in Fig.1-b. Notice that the mid-gap band is flat, i.e.,
it does not disperse, along $(0,\pi)-(\pi,\pi)$ and $(0,0)-(\pi,0)$ 
indicating that the holes are static in the horizontal direction (they
only can move vertically along the stripe). The corresponding schematic 
density of states is shown in Fig.1-c.

\begin{figure}[thbp]
\centerline{\psfig{figure=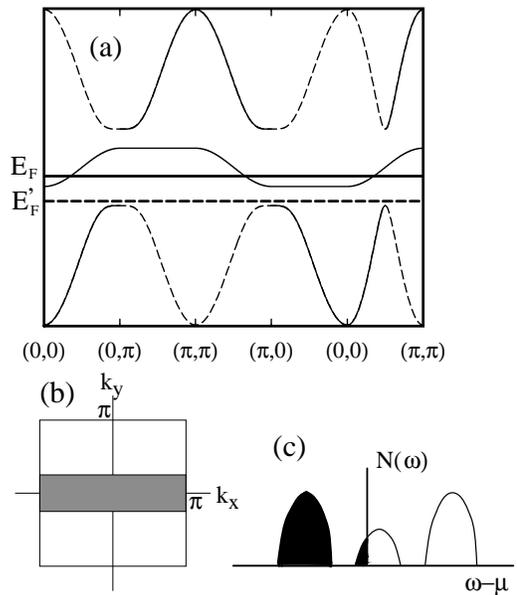,height=8cm}}
\caption{(a) Schematic band structure along the path
$(0,0)-(0,\pi)-(\pi,\pi)-(\pi,0)-(0,0)-(\pi,\pi)$ in the FBZ, for an 
AF system with metallic quarter-filled 
vertical stripes with $\langle n \rangle=0.875$. $E_F$ indicates the
position of the Fermi energy and shadow bands
are indicated by a dashed
line. $E'_F$
indicates the position of the Fermi energy in the case of empty 
($\langle n \rangle=0$) stripes.
(b) Fermi surface corresponding to the band structure shown in (a). (c)
The corresponding density of states of Fig.1-a.}
\end{figure}

This naive picture also shows that if
the hole density of the stripes were 1 rather than 0.5 the system would
be insulating since, as shown in Fig.1-a, the Fermi energy, indicated
by $E'_F$, would be inside a 
gap and all the stripe states would be empty. When typical mean-field
calculations allowing for the possibility of vertical stripes are
performed on the Hubbard model, the stable solution is qualitatively 
similar to the
one shown in Fig.1-a, with Fermi energy $E'_F$, 
indicating an insulating ground-state.\cite{Zaanen}
The addition of diagonal hopping terms distorts the
bands allowing the stabilization of metallic stripes.\cite{Ichioka}

Although the mean-field results mentioned above are very instructive, it
is not clear that they represent the true ground-state of the system.
In addition, 
numerical studies are very difficult to perform directly in the Hubbard model.
However, we have shown that the easier-to-study spin-fermion model
captures many of its properties.\cite{Charlie} We have also observed
that the addition of negative diagonal hopping parameters to the 
spin-fermion model with absolute 
values larger than $0.05t$ destabilize the 
striped ground-state on a square lattice.\cite{foot} 
This is the reason why, in spite of the encouraging mean-field
results of Ref.\cite{Ichioka}, we will here not show data for finite
diagonal hoppings.

\begin{figure}[thbp]
\centerline{\psfig{figure=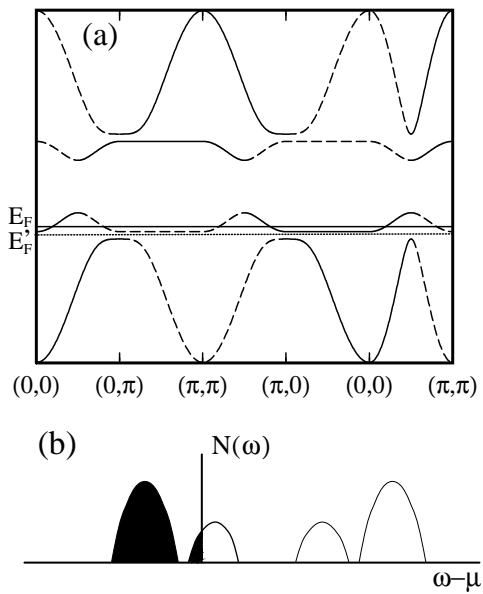,height=8cm}}
\caption{(a) Schematic band structure along the path
$(0,0)-(0,\pi)-(\pi,\pi)-(\pi,0)-(0,0)-(\pi,\pi)$ in the FBZ,
for an AF system with a splitted
mid-gap band
associated to metallic quarter-filled vertical 
stripes with $\langle n \rangle=0.875$. $E_F$ indicates the
position of the Fermi energy and shadow bands are indicated by a dashed
line. $E'_F$
indicates the position of the Fermi energy in the case of empty stripes.
(b) The corresponding density of states.}
\end{figure}

In order to compare the schematic results of Fig.1-a with 
those of the spin-fermion model it
has to be considered that antiferromagnetic
models tend to have symmetrical band structures below and above the
antiferromagnetic gap. Just as the conduction and valence band
described above are caused by an AF splitting, we should expect that the
mid-gap band will itself be splitted in the same way. In our schematic
model the energy of the splitted
mid-gap band is given by $E^{\pm}_m=\pm\sqrt{(t'
cos k_y)^2+\Delta'^2}$, where $2\Delta'$ is the gap between the two
bands. 
The expected band structure, using $\Delta'=0.5$ and $t'=0.7$, is 
schematically shown in Fig.2-a. This
indicates that, for vertical stripes, mid-gap spectral weight will
appear close to the valence band around $(\pi,0)$, while only shadow,
and thus much weaker, spectral
weight will appear around $(0,\pi)$. The observed mid-gap band should not
disperse from $(0,0)$ to $(\pi,0)$, but it will disperse from $(\pi,0)$
to $(\pi,\pi)$. It is also clear that very little mid-gap spectral weight
will be observed along the diagonal direction for $k$ larger than
$\pi/2$
since the shadow bands are not intense.
In this naive picture, the Fermi surface will be similar to the one
shown in Fig.1-b if $E_F$, in Fig.2-a, is the Fermi energy.\cite{foot1} 
As before,
no FS will be observed if the Fermi energy is $E'_F$ in Fig.2-a. The
DOS, on the other hand, will look qualitatively different to the one
displayed in Fig.1-c. Two mid-gap
bands will appear, instead of one, as shown in Fig.2-b. 

\begin{figure}[thbp]
\centerline{\psfig{figure=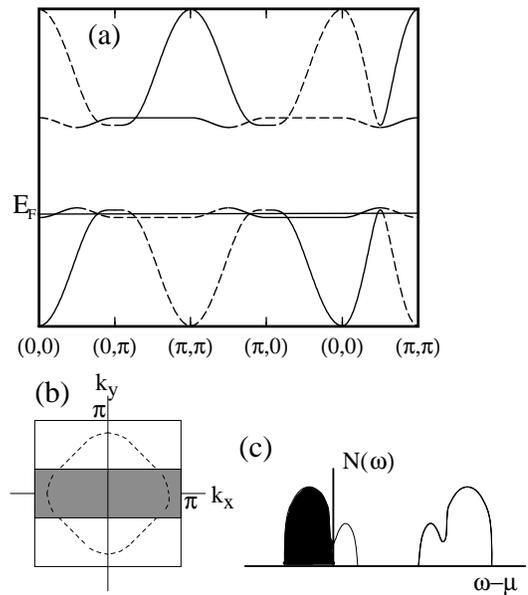,height=8cm}}
\caption{(a) Schematic band structure along the path
$(0,0)-(0,\pi)-(\pi,\pi)-(\pi,0)-(0,0)-(\pi,\pi)$ in the FBZ, 
for an AF system with overlapping
mid-gap and valence (conduction) bands
associated to metallic 
vertical stripes with overall $\langle n \rangle=0.875$. $E_F$ indicates the
position of the Fermi energy and shadow bands are indicated by a dashed
line. 
(b) Fermi surface corresponding to the band structure shown in (a). (c)
The corresponding density of states.}
\end{figure}

A problem with the naive schematic picture that we just discussed is
that, as it can be observed in Fig.2-b, the density of states (DOS) does 
not have a pseudogap at the chemical potential which is a feature
experimentally observed in the cuprates. In our schematic scenario, 
a pseudogap would be observed if the mid-gap states and the valence band 
overlap with each other. The resulting band structure is shown in
Fig.3-a. In this case, though, the Fermi surface would 
have an electron-like contribution coming from the valence band (doted line in
Fig.3-b), superimposed
to a FS qualitatively similar to the one shown in Fig.1-b due to the
partial filling of the mid-gap band, denoted by the full lines in Fig.3-b. 
The resulting DOS, with a pseudogap at the chemical potential, is shown in
Fig.3-c. We would like to point out that although here we have presented the
simplest case of band overlaping, it is also possible that the two
bands would merge together in which case a single FS would be observed.

\section{Results}

\subsection{Half-filling}

\begin{figure}[thbp]
\centerline{\psfig{figure=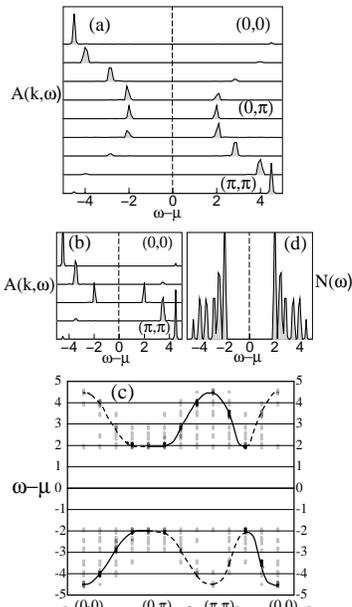,height=8cm}}
\caption{Spectral function for the undoped spin-fermion model on an $8
\times 8$ lattice as a function of $\omega-\mu$. (a) Along the path
$(0,0)-(0,\pi)-(\pi,\pi)$; (b) along the path
$(0,0)-(\pi,\pi)$; 
(c) the band structure obtained from (a) and (b). The lines indicating
the main bands are to guide the eye. Dashed lines
indicate shadow bands. Darker points indicate higher spectral weight; 
(d) the density
of states, $N(\omega)$ vs $\omega-\mu$.}
\end{figure}

Let us consider now the similarities and differences between the
schematic results described in the previous section and those arising
from the
spin-fermion model.
In Fig.4-a,b the spectral function along the main directions in the FBZ
is shown for the undoped $\langle n \rangle=1$ 
spin-fermion model on an $8\times 8$ cluster. 
The dispersion of the main
peaks in $A(k,\omega)$ gives rise to the band structure presented in
Fig.4-c, in which a valence and a conduction band separated by the AF gap
appear. The shadow bands associated with the AF order are indicated by
the dashed line in Fig.4-c. The ground-state is invariant under 
rotations in $\pi/2$, i.e., 
the features along the direction $(0,0)-(0,\pi)$ are identical
to those along $(0,0)-(\pi,0)$. The DOS showing the valence
and conduction bands separated by the AF gap is displayed in
Fig.4-d. The chemical potential is in the middle of the gap indicating that
the system is an insulator. The results shown here are to be expected
for an AF insulator, and they demonstrate that the spin-fermion model with
classical localized spins properly reproduces the main physics of such a 
state.

\subsection{${\bf\langle n \rangle=0.935}$}

In the MC simulations of the spin-fermion model,
holes added to the system tend to align forming horizontal or vertical
stripes, as discussed in previous literature.\cite{Charlie} 
Studying $8 \times 8$ and $12 \times 12$ systems it was observed that
for ${\rm x}$ less than $1/L$ holes doped ($L$ is the side of the square
cluster) the system remains insulator.
As an example, in Fig.5 we present $A(k,\omega)$ along the
main directions in the FBZ for $\langle n \rangle=0.937$, i.e., when 4
holes are introduced on the $8 \times 8$ system.

\begin{figure}[thbp]
\centerline{\psfig{figure=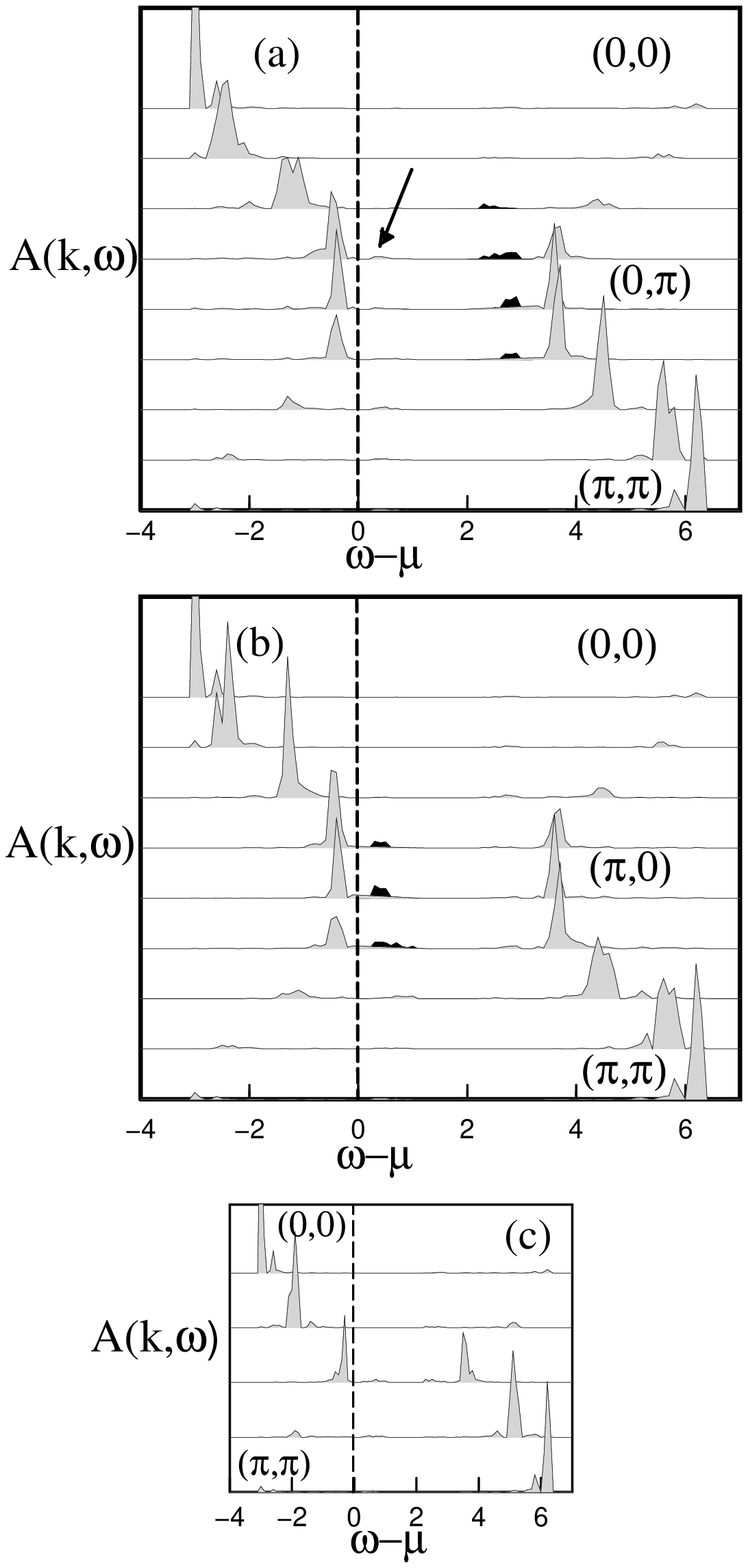,height=10cm}}
\caption{Spectral function for the spin-fermion model with
$\langle n \rangle=0.937$ on an $8
\times 8$ lattice, with holes aligned along the vertical axis forming a
straight segment. The mid-gap states are shaded in black. (a) Along the path
$(0,0)-(0,\pi)-(\pi,\pi)$. The arrow indicates the mid-gap shadow 
states with nearly negligible weight; 
(b) along the path
$(0,0)-(\pi,0)-(\pi,\pi)$; c) along the path
$(0,0)-(\pi,\pi)$.}
\end{figure}

\begin{figure}[thbp]
\centerline{\psfig{figure=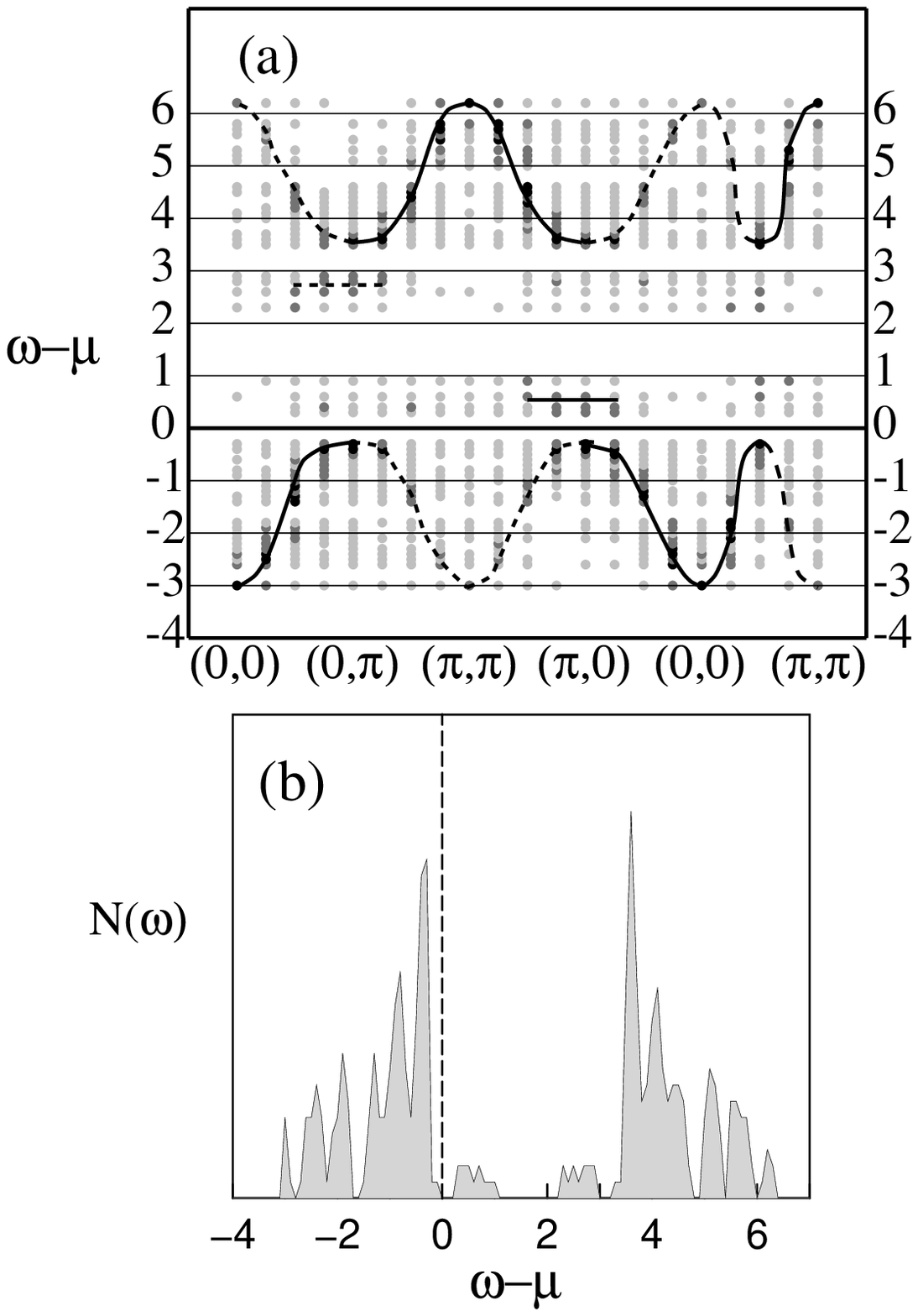,height=10cm}}
\caption{Band structure for the spin-fermion model with
$\langle n \rangle=0.937$ on an $8
\times 8$ lattice obtained from the data in Fig.5. (a) Along the path
$(0,0)-(0,\pi)-(\pi,\pi)-(\pi,0)-(0,0)-(\pi,\pi)$ in the FBZ. 
The valence, conduction, and mid-gaps bands are indicated with lines to
guide the eye. The dashed
lines connect points with very small spectral weights which define
shadow bands. (b) 
 The density of states, $N(\omega)$ vs $\omega-\mu$.}
\end{figure}

In Fig.5-a,b it can be seen 
that spectral weight is
transferred from the valence and conduction bands to the middle of the
gap, creating two mid-gap bands, shaded in black in Fig.5-a,b, 
similarly to the second schematic model described in section III. 
In the results shown here, the 4 doped holes are aligned vertically and, 
as expected, the spectral weight
associated with the holes
comes from the valence band, and it is located around $(\pi,0)$ in
momentum space (Fig.5-b). A redistribution of spectral weight, shaded in
black in Fig.5-a, is also 
observed close to the conduction band around $(0,\pi)$.
Only negligible shadow weight, indicated with an arrow in Fig.5-a, 
appears close to the valence band around $(0,\pi)$. 
It has been checked that, as expected, 
when the holes are aligned horizontally the states close to the
valence [conduction] band are located close to $(0,\pi)$ [$(\pi,0)$].
Also in agreement with the schematic picture, we observe that the lower
mid-gap states disperse along $(\pi,0)-(\pi,\pi)$, but not along
$(0,0)-(\pi,0)$. A somewhat unexpected result though is that the low mid-gap
spectral weight around $(0,0)$ is negligible, and the same occurs along
the diagonal direction. Thus, only states with momentum close to
$(\pi,0)$ and $(0,\pi)$ contribute appreciably to the mid-gap bands in the
spin-fermion model, in contrast with the schematic model of the previous
subsection in which all the
momenta participated. 

The resulting band structure is shown in 
Fig.6-a  and the DOS is presented in Fig.6-b. This band structure
is reminiscent of the one described in Fig.2 for vertical insulating 
1D stripes in an antiferromagnetic background.
The chemical potential
is in a gap between the lower mid-gap band, associated to the
doped holes in the stripes, and the valence band. The state is insulator
since the mid-gap bands are above the chemical potential. ARPES results 
for LSCO \cite{Ino} also detect an insulating ground-state at low doping.

The symmetry under $\pi/2$ rotations is
clearly broken due to the vertical stripe of holes. However, independent 
MC simulations, using different random initial spin configurations, 
provide outputs with vertical or horizontal stripes 
indicating that the two states are degenerate. 

As more holes are added to the system, in the insulating regime, spectral
weight is transferred from the valence and conduction bands to the
mid-gap states.

\subsection{${\bf\langle n \rangle=0.875}$}

\begin{figure}[thbp]
\centerline{\psfig{figure=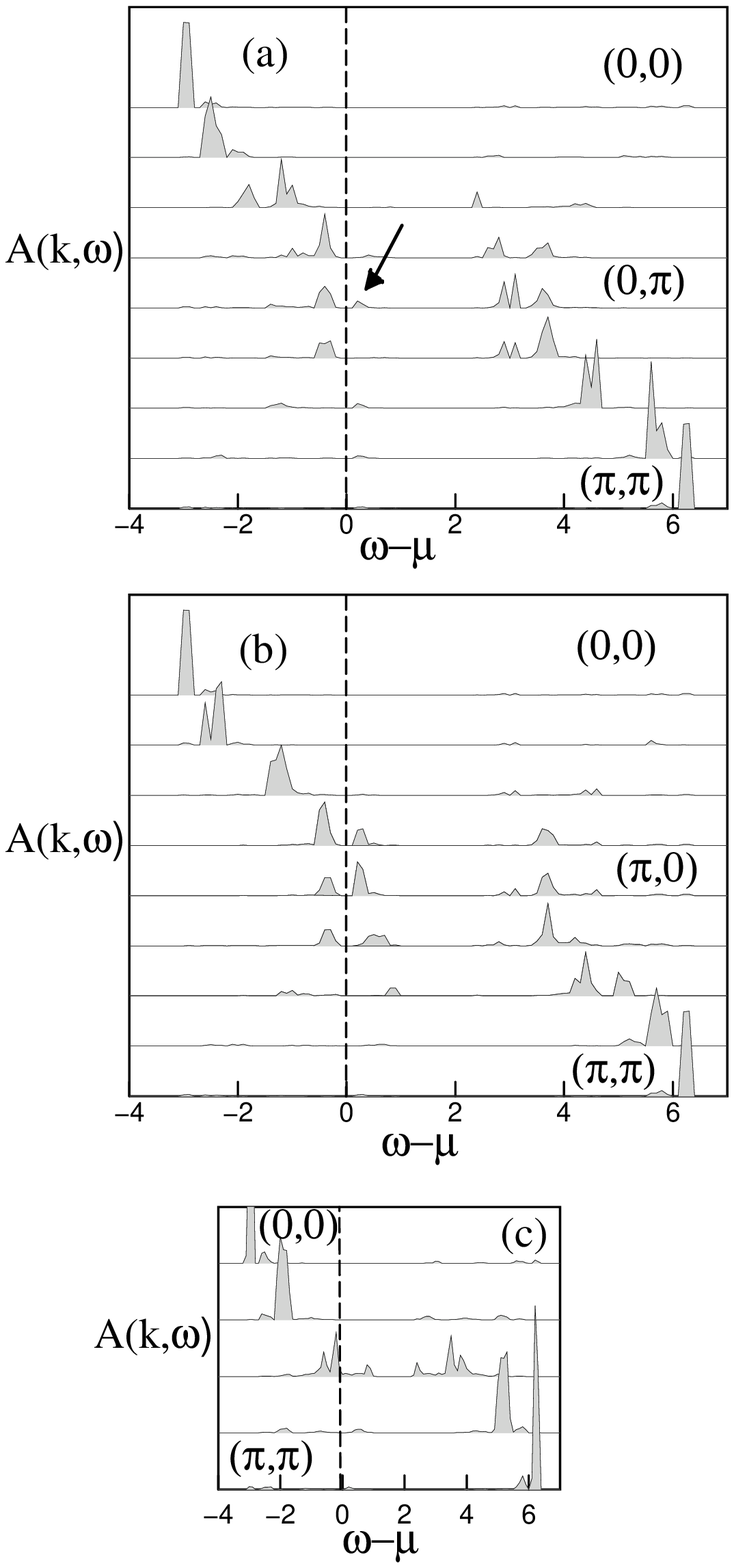,height=10cm}}
\caption{Spectral function for the spin-fermion model with
$\langle n \rangle=0.875$ on an $8
\times 8$ lattice. (a) Along the path
$(0,0)-(0,\pi)-(\pi,\pi)$. The arrow indicates the mid-gap shadow weight; 
(b) along the path $(0,0)-(\pi,0)-(\pi,\pi)$; (c) along the path
$(0,0)-(\pi,\pi)$.}
\end{figure}

At $\langle n \rangle=0.875$, which corresponds to 8 holes on the $8 \times
8$ size lattice, one stripe is stabilized in the ground-state. A very
important issue is whether the system is insulator, as in the early mean-field
results\cite{Zaanen}, or metallic as suggested by the experimental data.

The electronic density of the stripes observed in the spin-fermion model 
is $\sim 0.5$, 
\cite{Charlie} in agreement
with the experiments, but there is a slight reduction of the
density of all the sites of the lattice indicating that, on average, 4
holes are located at the stripe, while the other 4 are spread in the
system. We will discuss this point in more detail below. Notice that this
is different from the picture emerging from experiments performed
in Nd-doped LSCO\cite{Tran} in which two rather than one stripe would
be expected to be observed in the situation that we are discussing, since
the holes, are assumed to be located exclusively in the quarter-filled stripes 
and not in the background. 
This could be a drawback of this model since the
incommensurability pattern for LSCO is similar to the Nd-doped material
mentioned above.\cite{yamada} 
However, it is not clear that all the cuprates follow
the same pattern.\cite{Dai} The spectral functions in the FBZ are shown
in Fig.7 and they correspond to the case in which the stripe is
vertical. The same results, exchanging $k_x$ with $k_y$, were obtained for
the case of an horizontal stripe. In Fig.7-a it can be observed that the
lower mid-gap states only have very small shadow weight, indicated with
an arrow, around $(0,\pi)$, 
as expected from the schematic model, and the valence band remains below
the chemical potential. Thus, it appears
that there is no FS along this direction. 

\begin{figure}[thbp]
\centerline{\psfig{figure=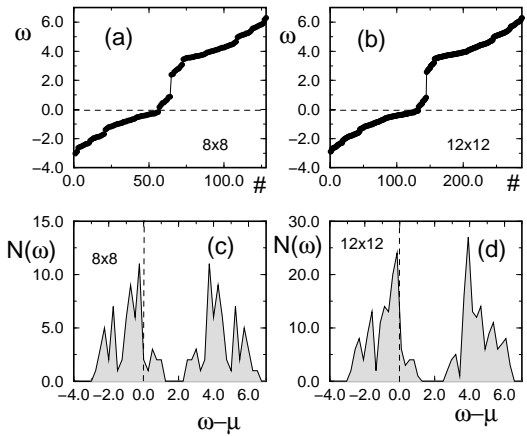,width=8cm}}
\caption{(a) Eigenvalue distribution for a snapshot configuration on 
an $8 \times 8$ cluster with a vertical stripe; (b) same as (a) but for a
$12 \times 12$ cluster; (c) density of states obtained from (a);
(d) density of states obtained from (b).}  
\end{figure}

Along the
$(0,0)-(\pi,0)-(\pi,\pi)$ direction, depicted in Fig.7-b, on the other
hand, the valence band and the lower mid-gap band are very close to each
other.
On the $8 \times 8$ system (Fig.7-b), there seems to exist
a very small gap in between the valence and the mid-gap band which would 
indicate insulating behavior. However, we need to establish whether this
gap is real or an artifact of the small lattice that we are studying. 
To explore this issue we studied the
eigenvalue distribution of the Hamiltonian defined in Eq.(1) for several
configurations of the spin variables (snapshots) whose ground-state had one 
stripe on $8 \times 8$
(Fig.8-a) and $12 \times 12$ (Fig.8-b) systems. We observed that the 
eigenvalue distribution
changes very little for the different snapshots. On the $8 \times 8$ system
a very small gap in the eigenvalues distribution was observed at the
chemical potential indicated with a dashed line in Fig.8-a.
As expected, there are 8
eigenstates forming the lower mid-gap states associated with the number
of holes in the system. 

\begin{figure}[thbp]
\centerline{\psfig{figure=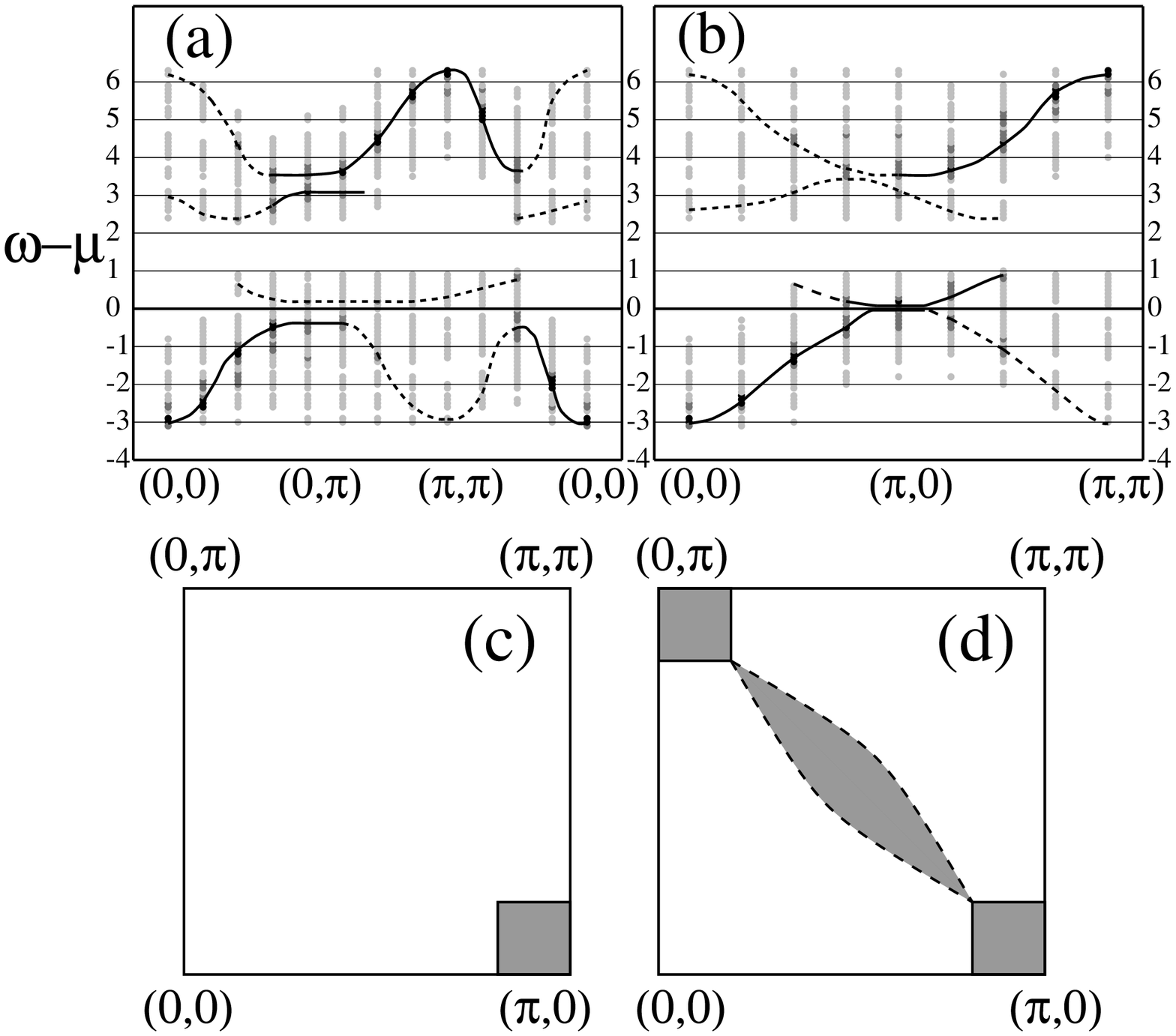,height=8cm}}
\caption{Band structure for the spin-fermion model with
$\langle n \rangle=0.875$ on an $8
\times 8$ lattice obtained from the data in Fig.5. (a)
Along the path
$(0,0)-(0,\pi)-(\pi,\pi)-(0,0)$ in the FBZ; (b) along the path
$(0,0)-(\pi,0)-(\pi,\pi)$; 
(c) possible FS obtained from the data
in Fig.7; (d) possible FS
obtained by combining the data for vertical and horizontal stripes.
The solid lines indicate an actual FS, while the dashed line indicates a
gapped region or very weak FS.}
\end{figure}

For the $12 \times 12$ system (Fig.8-b)
the gap between the lower band and
mid-gap states is smaller. The number of
eigenstates per energy interval provides the density of states for the
snapshot. An histogram of the eigenvalue distribution provides an
estimation of the DOS. These histograms, using an energy interval
equal to 0.04$t$, are shown in Fig.8-c for the
$8 \times 8$ system, and in Fig.8-d for the $12 \times 12$ cluster. In
both cases the chemical potential is very close to a pseudogap in the
DOS located between the valence and lower mid-gap band, as 
experimentally observed in the cuprates.\cite{Sato} 

Thus, combining the results in Fig.7 and Fig.8 we infer that, in this
case, both the valence and the low mid-gap band merge together and
will contribute to the
Fermi surface. It is not possible to detect the position of the FS
precisely due to the flatness of the overlapped bands close to $(\pi,0)$.
Our data are consistent with either a FS at  
$(\pi-\delta,0)$ or at $(\pi,\delta)$, where $0 \leq\delta<3\pi/4$. 

Notice that the dispersive behavior of the mid-gap states is the one
expected for vertical stripes in the schematic system, although the
dispersion is small.

A similar analysis looking at Fig.7-c shows that a FS 
along the diagonal of the FBZ
i.e., from $(0,0)$ to $(\pi,\pi)$ does not exist. The
valence band quasi-particle peak, which is well defined at $(0,0)$,
becomes incoherent just below the chemical potential for momentum
$(\pi/2,\pi/2)$, and it is almost touching the Fermi energy.
The mid-gap states, as expected, have only shadow spectral weight. 

The band structure arising from Fig.7 is presented in Fig.9-a, b and the
possible 
FS due to the vertical stripe is shown in Fig.9-c. The shadowed region
indicates its possible location and the solid lines running from
$(7\pi/8,0)$ to $(7\pi/8,\pi/8)$ and $(\pi,\pi/8)$ to $(7\pi/8,\pi/8)$
denote its two approximate extreme positions. 

In Fig.9-d we show the possible FS obtained as a superposition of the 
results for horizontal
and vertical stripes. It is an incomplete FS which could close either
around $(0,0)$ or around $(\pi,\pi)$ (solid
line in Fig.9-d). We could not determine if the surface is close or 
open but away
from $(\pi,0)$ and $(0,\pi)$ the spectral weight is weak and incoherent.
The possible extreme positions 
of this weaker FS (or gap) is indicated with dashed lines
in Fig.9-d.  
The FS closing around $(\pi,\pi)$ has been observed in underdoped
LSCO \cite{Ino} and Bi2212 \cite{Feng,Fretwell}.
The observation of a FS closing around
$(0,0)$ is controversial. 
According to Ref. \cite{Feng} it appears
in Bi2212 at 32.3 eV photon energy, while Ref. \cite{Fretwell}
interprets similar data as indicating no changes in the shape of the
FS. 

Our results are also in agreement with the LSCO data
showing a weak or no FS along the diagonal in the underdoped 
regime.\cite{Ino}
ARPES data seem to indicate a FS along the diagonal direction in Bi2212 
\cite{Feng}, but we believe that inverse ARPES data would be necessary to 
rule out the existence of a gap. In our results the existence of the gap
becomes clear after observing the behavior of the spectral weight in the
valence band and the mid-gap band above $\mu$.

\subsection{${\bf\langle n \rangle=0.80}$}

Next we will study how the band structure evolves as the system is doped
further away from half-filling.
At $\sim 20\%$ doping 
it is possible to determine that the FS closes around $(0,0)$. 
The symmetry under rotations is still spontaneously broken, but the
potential barrier separating vertical and horizontal stripes appears to
be lower than for lighter doping. As a result, some features due to
horizontal configurations appear in the spectral functions presented
here which corresponds to a ground-state with mostly vertical charge
inhomogeneities. 

The
spectral functions along the main directions in the FBZ are shown if
Fig.10-a, b and c, and the corresponding band structure is in
Fig.10-d and e. As in the previous case,
a gap rather than a FS is observed close to $(0,\pi)$ (Fig.10-a and d).
\cite{peak} Close to $(\pi,0)$ the 
valence and mid-gap bands overlap with each other 
where the FS seems to be located (Fig.10-b and e). The FS
along the diagonal is still inexistent or very weak since the spectral
weight at $(\pi/2,\pi/2)$ is incoherent (Fig.10-c).

\begin{figure}[thbp]
\centerline{\psfig{figure=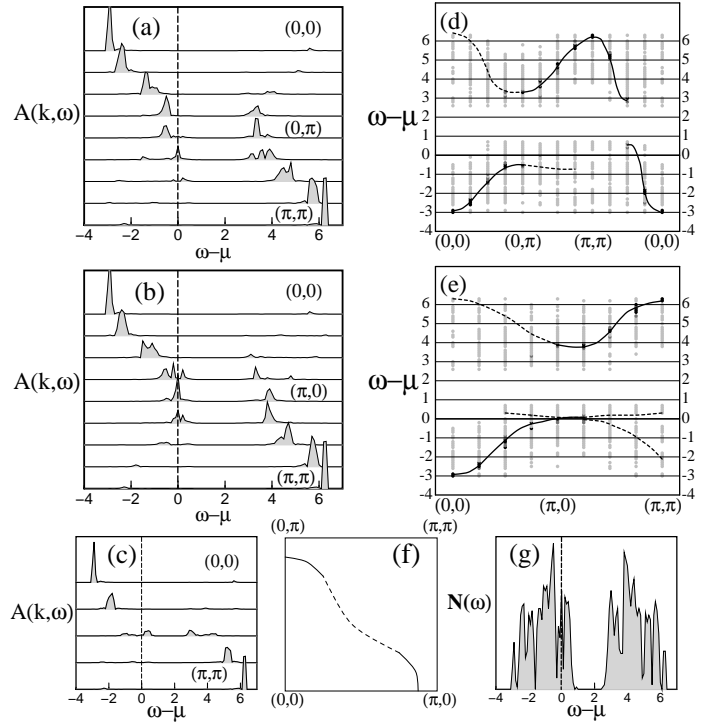,height=10cm}}
\caption{Spectral functions for the spin-fermion model with
$\langle n \rangle=0.81$ on an $8
\times 8$ lattice. (a)
Along the path
$(0,0)-(0,\pi)-(\pi,\pi)$ in the FBZ; (b) along the path
$(0,0)-(\pi,0)-(\pi,\pi)$; (c) along the path $(0,0)-(\pi,\pi)$; 
(d) band structure along the path $(0,0)-(0,\pi)-(\pi,\pi)-(0,0)$;
(e) along the path $(0,0)-(\pi,0)-(\pi,\pi)$;
(f) FS obtained by combining the data for vertical and horizontal
stripes. The dashed line indicates a very weak FS; (g) The density of states.}
\end{figure}

The shape of the FS for
the combined ground-state (with vertical and horizontal
incommensurability) is shown in Fig.10-f. This FS was obtained by
analyzing $A(k,\omega)$ for all the available momenta in the FBZ.
In Fig.10-g we present the DOS
and it can be seen that the chemical potential is in a pseudogap.

\subsection{${\bf\langle n \rangle=0.75}$}

When 16 holes are introduced
in the system two parallel stripes are stabilized. 
The system is metallic and the 
FS closes around (0,0). The spectral functions along
$(0,0)-(0,\pi)-(\pi,\pi)$ are shown in Fig.11-a. It is clear that the
valence band does not cross the chemical potential, and only shadow
weight appears in the lower mid-gap band. Along $(0,0)-(\pi,0)-(\pi,\pi)$,
presented in Fig.11-b, the valence and low gap bands overlap and cross
the chemical potential at $k\sim(3\pi/4,0)$. The FS in the
diagonal direction is weak because the spectral weight that crosses
$\omega=\mu $ is incoherent as it can be seen in Fig.11-c.  
The band structure corresponding to the case in
which the stripes are vertical is shown in Fig.11-d and e.
The superposition of the two ground-states (vertical
and horizontal stripes) will give raise to a FS 
which closes around (0,0) as shown in Fig.11-f. Clearly the chemical
potential is located in a pseudogap in the density of states, as it can
be seen in Fig.11-g.  

\begin{figure}[thbp]
\centerline{\psfig{figure=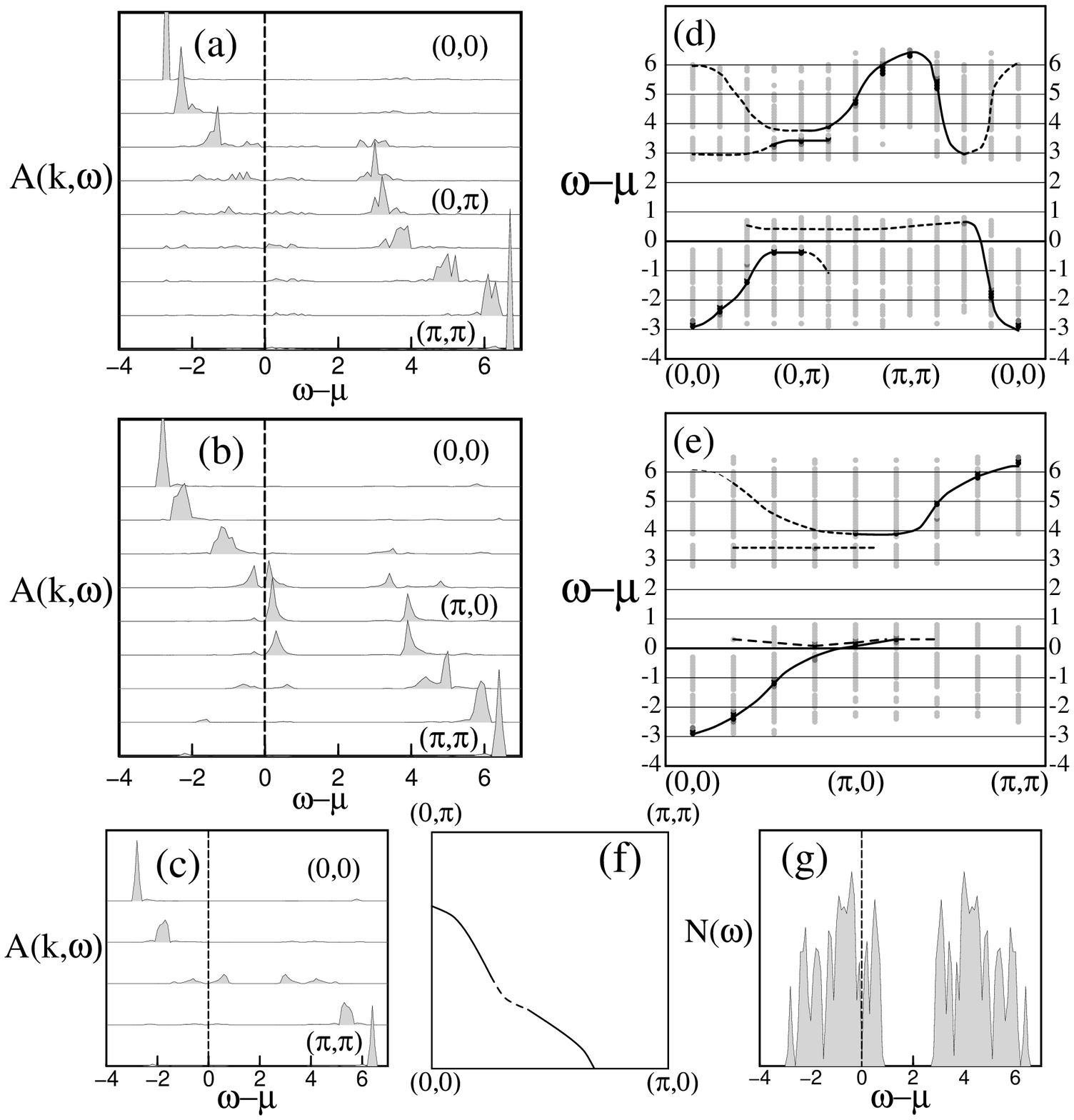,height=10cm}}
\caption{Spectral functions for the spin-fermion model with
$\langle n \rangle=0.75$ on an $8
\times 8$ lattice. (a)
Along the path
$(0,0)-(0,\pi)-(\pi,\pi)$ in the FBZ; (b) along the path
$(0,0)-(\pi,0)-(\pi,\pi)$; (c) along the path $(0,0)-(\pi,\pi)$; 
(d) band structure along the path $(0,0)-(0,\pi)-(\pi,\pi)-(0,0)$;
(e)along the path $(0,0)-(\pi,0)-(\pi,\pi)$
(f) FS obtained by combining the data for vertical and horizontal
stripes. The dashed line indicates a very weak FS; (g) The density of states.}
\end{figure}

\subsection{${\bf\langle n \rangle=0.625}$}

With increasing doping the area of the FS is reduced. For $\langle n
\rangle=0.625$ the symmetry under rotations is still broken. The
magnetic and charge incommensurability still form vertical or horizontal 
patterns. The spectral functions for the vertical pattern are shown in
Fig.12-a, b and c. In Fig.12-a it can be seen that no FS
is observed along $(0,0)-(0,\pi)-(\pi,\pi)$. Fig.12-b shows that
the mid-gap and valence band which had merged together
cross $\mu$ below $(3\pi/4,0)$. In the diagonal direction just below
$(\pi/2,\pi/2)$ incoherent weight crosses the chemical potential
determining a weak FS (Fig.12-c). The corresponding band structures are
presented in Fig.12-d and e and the FS is shown in Fig.12-f. 
The chemical potential is in a pseudogap as shown in Fig. 12-g.

\begin{figure}[thbp]
\centerline{\psfig{figure=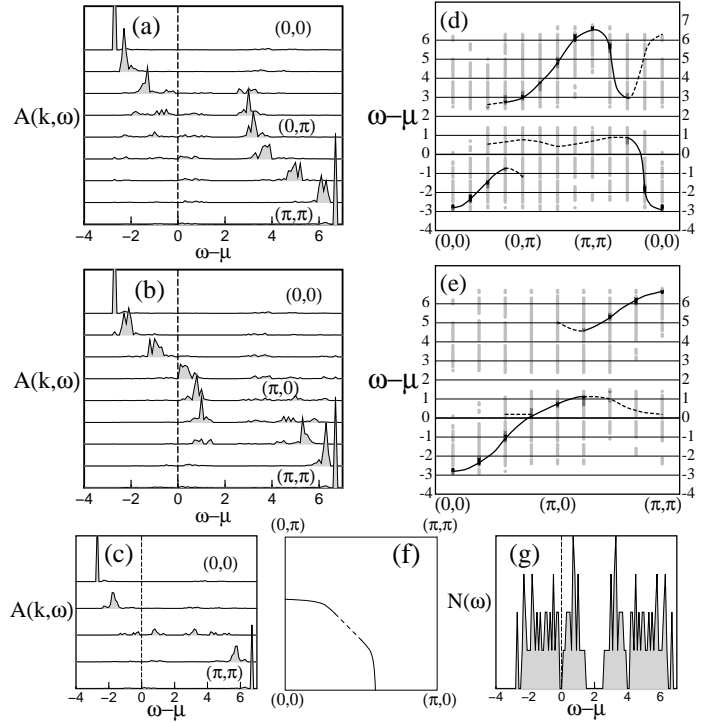,height=10cm}}
\caption{Spectral functions for the spin-fermion model with
$\langle n \rangle=0.625$ on an $8
\times 8$ lattice. (a)
Along the path
$(0,0)-(0,\pi)-(\pi,\pi)$ in the FBZ; (b) along the path
$(0,0)-(\pi,0)-(\pi,\pi)$; (c) along the path $(0,0)-(\pi,\pi)$; 
(d) band structure along the path $(0,0)-(0,\pi)-(\pi,\pi)-(0,0)$;
(e)along the path $(0,0)-(\pi,0)-(\pi,\pi)$
(f) FS obtained by combining the data for vertical and horizontal
stripes. The dashed line indicates a very weak FS; (g) The density of
states.}
\end{figure}

\subsection{Analysis}

The previous results indicate that as
the electronic density varies from 1.0 to 0.625  a
clear change in the band structure of the system occurs.
Spectral weight for the momenta close to the chemical
potential is transfered from the valence band to the lower mid-gap band. 
As a result, the chemical potential always appears in a pseudogap
between the valence and the lower mid-gap bands.
The bottom of the valence band, located at $k=(0,0)$, remains at an
approximately constant
distance from the chemical potential. This is very different from the
mean-field studies \cite{Ichioka} in which a rigid band structure is
obtained and the chemical potential approaches the bottom of the valence
band as holes are added. We believe that the main difference between our
numerical results and mean-field is due to 
the fact that in the
mean-field analysis the proposed striped background is always an almost 
perfect antiferromagnet, which physically is not accurate.
Another important characteristic of the evolution of the band structure
with doping is that while the gap between the valence and lower mid-gap
band closes becoming a pseudogap, the gap between the valence and 
conduction band still exists even at 40\% doping. 

Another characteristic of the band structure in the metallic case is
that for vertical incommensurability, the quasiparticle at $(0,\pi)$ has
lower energy than at $(\pi,0)$ as it can be seen in Fig.9-a,b,
and in parts (d) and (e) of Figs.10, 11 and 12. If we consider a 2D
system of free electrons with horizontal hopping $t_x$ and vertical
hopping $t_y$, the energy as a function of momentum is given by
$\epsilon_k=-2(t_x cos(k_x)+t_y cos(k_y))$. It is clear that $\epsilon
(\pi,0)>\epsilon(0,\pi)$, if $t_x>t_y$. This may indicate that, in the
spin-fermion model, it is easier for the electrons to move in the
direction perpendicular to the stripes rather than along them.   

\section{The Pseudogap}

The results presented in the previous section show that, upon hole
doping, a pseudogap between the valence and the lower mid-gap
stripe-induced band
develops at the chemical potential in the density of states (see
Fig.8-c,d, and part (g) of Figs.10, 11 and 12). This is a feature that has
been observed experimentally both in the cuprates\cite{Sato} and in the
manganites \cite{dessau}, and it deserves to be understood in terms of
the spin-fermion model. 

The formation of a pseudogap at the chemical
potential in the manganites has been explained in previous studies
as arising from the coexistence of hole-poor
antiferromagnetic insulating regions and hole-rich ferromagnetic 
clusters.\cite{pseudo} In Ref.\cite{pseudo} it was concluded that
pseudogap behavior should be observed in any compound which is in a
mixed-phase regime. Although the spin-fermion model does not present phase
separation for the value of J studied in the present paper, it is clear     
that the ``striped'' ground state has mixed-phase characteristics, since the
stripes are richer in holes than the background. A schematic picture of
the hole distribution in the direction perpendicular to the stripes is
shown in Fig.13-a. The density of
holes ${\rm x}$ in the direction perpendicular to the stripes is 
clearly non-uniform, and
it is schematically shown 
in Fig.13-b. We can represent this result using an 
``effective potential'' for the holes which 
will be more negative at the stripes, as shown in Fig.13-c. The states
inside the wells generate the mid-gap bands along the direction parallel
to the stripes in the BZ, $(\pi,0)-(\pi,\pi)$ for vertical stripes. 
Clear quasi-particle peaks and maximum
dispersion are expected along this direction since it is easier for the
holes to move along the stripes. Along the direction perpendicular to
the stripes, from $(0,0)$ to $(\pi,0)$ for vertical stripes, 
the holes move through tunneling between the potential
wells or by fluctuations away from the stripe, effectively providing a
width $d$ to such stripe. This motion will generate a mid-gap band,
schematically shown in Fig.13-c. The electrons, on the other
hand, will move mostly across the stripes, as discussed in the previous
section and thus, a dispersive valence band would be expected.
As a result of the formation of the
mid-gap bands a pseudogap develops at the chemical potential in the 
DOS, as schematically shown in Fig.13-d.

\begin{figure}[thbp]
\centerline{\psfig{figure=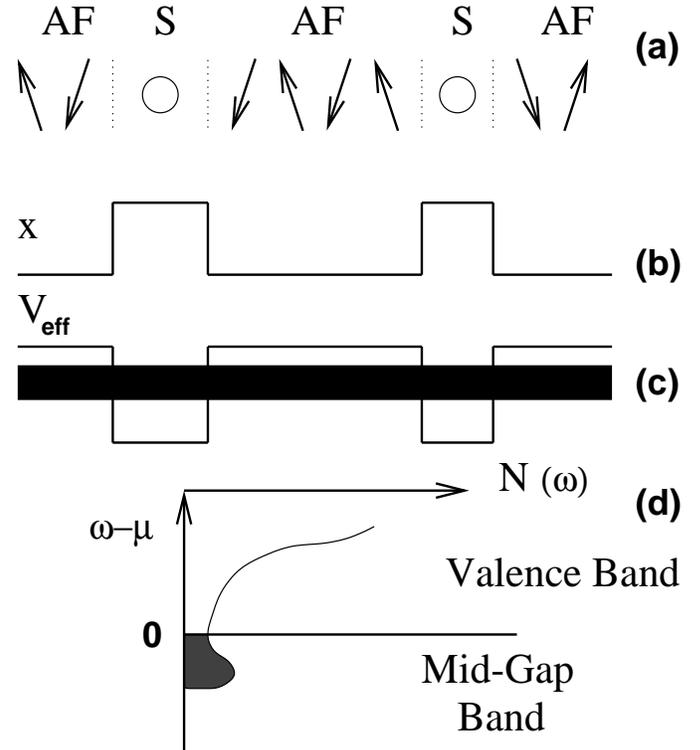,height=10cm}}
\caption{Schematic explanation of the pseudogap formation upon hole
doping in the spin-fermion model. In (a) a typical distribution of the
AF background and stripes (S) reach in holes along the
direction perpendicular to the stripes is shown. In (b), the
corresponding hole density $ x $ is sketched, showing that
the holes are mostly located in the stripes. In (c), the effective
potential felt by the holes is presented. The thick line indicates the
mid-gap band populated with holes which develops. The resulting DOS
with the chemical potential in the pseudogap is sketched in (d).}
\end{figure}

An important difference between the pseudogap in the spin-fermion model
and the one observed in models for manganites is that in the latter the
pseudogap is isotropic in momentum space. In the spin-fermion model, on
the other hand, the pseudogap is clearly observed close to
$(\pi,0)$, while a clear gap appears close to $(0,\pi)$ (for vertical
stripes) since the lower mid-gap band has only shadow weight here and
the valence band has a lower energy than at $(\pi,0)$.
This difference should be attributed to the
one-dimensional nature of the stripes in the spin-fermion model. The
incoherence of the spectral weight along the diagonal is, as mentioned
above,  due to the fact that diagonal hopping of electrons or holes is not
particularly favored.

\section{Conclusions}

Summarizing, we have studied the spectral functions of the spin-fermion
model, which has a ground-state in which added holes tend to form
vertical and horizontal 
stripes, using unbiased numerical techniques. We have observed that 
doped holes contribute to the formation of mid-gap bands by
modifying the valence and conduction bands associated to the 
insulator. The mid-gap bands arise from a dynamically generated 
effective potential which
produces a non-uniform distribution of charge in the ground-state.
In the metallic regime the lower mid-gap and valence band
overlap with each other, giving rise to a pseudogap in the density of
states at the chemical potential, as observed in experiments for the
cuprates. A pseudogap arising from inhomogeneities in the ground-state
has also been observed in previous investigations of models for manganites.

The ground-state of the spin-fermion model appears to change from 
an insulator to a conductor with 
an incomplete FS when a single stripe gets
stabilized, i.e., for a hole density $x=1/L$. 
In the conductor, electrons hop more easily in the direction
perpendicular to the stripes. Thus, though the striped state in the
spin-fermion model is not insulating, it appears that most of the
electronic hopping occurs in the direction perpendicular to the
stripes rather than along the stripes.
Close to 20\% doping the
FS  clearly closes around (0,0). These features are qualitatively
similar to those observed in ARPES experiments. The band structure is
not rigid, in disagreement with mean-field results, and it smoothly changes
from insulating to conductor. 
Another interesting feature is the observation of well defined
quasi-particle peaks in $A(k,\omega)$ which become incoherent as the
chemical potential is approached indicating that the FS is very weak or
it does not exist along the diagonal direction. Also close to $(0,\pi)$
[$(\pi,0)$] in the case of vertical [horizontal] stripes the chemical
potential is in a gap. These features are related to the
one-dimensionality of the stripes. A similar behavior of the valence and
mid-gap band has been observed using a cluster perturbation theory for
the $t-J$ model.\cite{eder}

Although part of our results disagree with some of the theoretical models
for stripes in the cuprates, we believe that this numerical study provides
unbiased information on the dynamical properties of charge inhomogeneous
ground states which will be relevant to understand the behavior of
materials whose inhomogeneous properties are just beginning to be
unveiled experimentally.

\section {Acknowledgements}

We would like to thank G. Sawatzky and E. Dagotto for their useful comments.
A.M. is supported by NSF under grant DMR-9814350.
Additional support is provided by the National High Magnetic Field Lab 
and MARTECH.


\begin{references}
\bibitem{foot} We are presently studying the model using ladders with
positive as well as negative hopping.
\end{references}
\end{document}